\documentclass{evn2004}
\usepackage{txfonts}
\begin{document}
   \title{EVN, MERLIN, and VLA Observations of NRAO530}

   \author{X. Y. Hong\inst{1}
          \and
          J. H. Zhao\inst{2}
          \and
          T. An\inst{1}
          \and
          D. R. Jiang\inst{1}
          \and
          W. H.~Wang\inst{1}
          \and W. X. Feng\inst{1}
          \and C. H. Sun\inst{1}}

   \institute{ Shanghai Astronomical Observatory, 80 Nandan Road, Shanghai
  200030, China         \and
              Harvard-Smithsonian CfA, 60 Garden St, MS 78, Cambridge, MA
              02138}

   \abstract{
We present images of NRAO530 observed with the EVN (VLBI) at 5
GHz, the MERLIN at 1 .6 and 5 GHz, and the VLA at 5 and 8 GHz
showing the complex morphology on scales from pc to kpc. The VLBI
image shows a core-jet structure indicating a somehow oscillation
trajectory on a scale of 30 mas, north to the strongest compact
component (core). A core-jet structure extended to several
hundreds mas at about  P.A. $-50 \degr$ and a distant component
located 11 arcsec west to the core are detected in both the MERLIN
and the VLA observations. An arched structure of significant
emission between the core and the distant component is also
revealed in both the MERLIN image at 1.6 cm and the VLA images at
8.4 and 5 GHz. The core component shows a flat spectrum with
$\alpha\sim -0.02$ ($S \propto \nu^{-\alpha}$) while
$\alpha\sim0.8$ for the distant component. The steep spectrum of
the distant component and the detection of the arched emission
suggests that the western distant component is a lobe or a
hot-spot powered by the nucleus of NRAO530. A patch of diffuse
emission, 12 arcsec nearly east (P.A. $70\degr$) to the core
component, is also observed with the VLA at 5 GHz, suggesting a
presence of a counter lobe in the source.

   }

   \maketitle
%

\section{Introduction}

The quasar NRAO530 is a well known optically violently variable
(OVV) extragalactic source with $m_{pg} \sim 18.5$ mag (Whelch and
Spinrad 1973) at a redshift of 0.902 (Junkkarinen 1984). It is an
object of intensive observations at all wavelengths from radio to
$\gamma$-ray. Weak polarization of the source was detected both at
optical and radio bands. NRAO530 was detected by ROSAT with
$1.84\times10^{-6}$~Jy at 1.3 keV (Brinkmann et al. 1994) and
identified as a $\gamma$-ray source by the Energetic Gamma Ray
Experiment Telescope (EGRET) with flux density of
$4.6\times10^{-11}$~Jy at 2.55~GeV (Fichtel et al. 1994; Thompson
et al. 1995). The detection of the source in high frequency
indicates that the evidence for relativistic bulk motion of the
emitting plasma exists in its nucleus. A Doppler factor of 5.2 of
NRAO530 was estimated by assuming that the particles and magnetic
field are in equipartition (Guuosa \& Daly 1996).

NRAO530 has been extensively monitored as a low frequency variable
source (e.g., Bondi et al. 1994). Three epochs early VLBI
observations at 1.7 GHz showed the structure was oriented in
north-south direction extended to 25 mas in P.A.$ -7\degr$ (Romney
et al. 1984, Bondi et al. 1994). The morphology of the maps shows
two-sided structures. Its complex structure variations are not
readily interpretable as angular expansion because the separation
of components and the component sizes appear unchanged (Bondi et
al. 1994).

Two epochs VLBI observations at 5 GHz showed a 4 mas slightly
curved jet of the source towards to around the North direction,
and no counter jet has been detected (Shen et al. 1997, and Hong
et al. 1999). A jet-like feature extending approximately 5 mas
from the core along P.A.$\sim 15\degr$ was seen on the VLBI image
at 8.5~GHz by Tingay et al. (1998).

The source underwent a dramatic radio outburst in 1995 with
amplitude higher than any found over 30 years of monitoring (Bower
et al. 1997). The creation of new components in the jet associated
with the outburst was revealed by their 86~GHz VLBI observation
with one component (C1, their notation) to a radio flare observed
in mid of 1994 and another component (C2) to the 1995 outburst.
The components moved to the southwest with apparent velocities of
7.4 and 7.9 $h^{-1}~c$, respectively, by assuming the brightest
component as the core (Bower et al. 1997).

The 22 and 43~GHz VLBA monitoring of the source has been made
during the flare in 1995 (5 epoch from 1994.45 to 1996.90) by
Jorstad et al. (2001).  Based on better time sampling, they
interpreted the evolution by assuming the south component as the
core and the brightness feature is the ejected component during
the flare. This results shows the structure and motion are both
oriented roughly to the north with apparent velocities in the
range of 7 to 29$~h^{-1}~c$, with a stationary feature to the
north at the distance of $\sim$ 1.4 mas from the core (Jorstad et
al. 2001).

VLA observation of the source at 1.4 GHz showed an unresolved core
and a second unresolved component 11$\arcsec$ around in P.A.
$-90\degr$ (Perley 1982) without evidence for a connection between
the two (Romney et al. 1984).

In this paper, we focus on the extended radio emission of the
source at the kpc scale and the relation between the pc and kpc
structures with EVN (European VLBI Network), MERLIN, and VLA
observations at multi-frequencies.

\section{The observations and data reduction}

   \begin{figure}
   \centering
   \vspace{80pt}
   \includegraphics{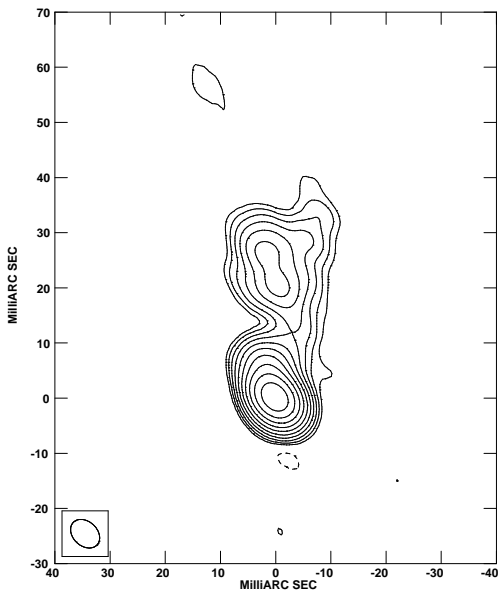}
 \vspace*{5.5cm}
   \caption{EVN image of NRAO530 at 5 GHz. Peak flux density is 2.7 Jy/beam; the $r.m.s.$
 noise value is  0.5 mJy/beam; the lowest contours is 3 mJy/beam and
 contour levels increase by a factor of 2. FWHM:
 5.9$\times4.4$ mas; P.A. $46.6\degr$ }
    \end{figure}

\begin{table}\centering
\caption{Spectral index ($S\propto\nu^{-\alpha}$) of  gaussian
components} \vspace{-2mm}
\begin{tabular}{ccccc}\hline
\\{\smallskip}
 Comp.   &$\alpha^{5(MERLIN)}_{1.6(MERLIN)}$&
$\alpha^{8(VLA)}_{5(VLA)}$&$\alpha^{8(VLA)}_{1.6(MERLIN)}$&$\alpha^{*}$\\\hline
\\
{\smallskip}
 Core       &  0.08 &$-0.5$  &$-0.06$  &$-0.02$\\ {\smallskip}
 West lobe  &  1.0  & 1.0    &0.8      &0.8    \\\hline
\end{tabular}\\
{\smallskip} note: $\alpha^{*}$ is calculated with least squares
between 1.6 and 8 GHz.
\end{table}

A full-track 7-hour EVN + MERLIN observation (EH004) of the blazar
NRAO530 was carried out at 5 GHz on February 19, 1997. The EVN
array comprised Effelsberg, Shanghai, Cambridge, Jodrell Bank
(Mark~2), Medicina, Noto, Onsala, Urumqi, WSRT, and Torun. The
data were acquired with the Mk\,III VLBI recording system in Mode
B with an effective bandwidth of 28 MHz and correlated at the
Max-Plank Institut f\"ur Radioastronomie in Bonn. The EVN data
were calibrated and fringe-fitted using the NRAO (Astronomical
Image Processing System) AIPS package.  The initial amplitude
calibration was accomplished using the system temperature
measurements made during the observations and the priori station
gain curves.

The MERLIN array of the EVN + MERLIN observation comprised  6
antennas (Defford, Cambridge, Knockin, Darnhall, Mark~2, and
Tabley). The central frequency was 4.994 MHz and the observing
bandwidth was 14 MHz.

An archive MERLIN data of NRAO530 at 1.6~GHz is available for our
study. NRAO530 was observed as phase reference source with
snapshot mode for total 2 hours on source with 46 scans over a
period of 8 hours. The observation was carried out on May 8, 1998.
The MERLIN array consists of 7 antennas (the same as above plus
Lovell).

The quasar NRAO530 was observed with the VLA on Nov. 27, 1992 at
8.4 GHz and on June 17, 2003 (AZ143) at 4.8 GHz as a
calibrator. The observations were carried out in snapshot mode.
and used the standard VLA recording configuration (50~MHz
bandwidth).

Post-processing including editing, phase and amplitude
self-calibration, and imaging of the data were conducted in the
AIPS package. Natural weighting was used for imaging to obtain the
extended emission.


\section{Results}

\subsection{VLBI image}

The 5~GHz VLBI image displayed a jet  to the North with a series
of components on the scale of 30 mas with a somewhat oscillating
trajectory (Fig. 1). The jet first moved in P.A. $15\degr$  for
about 5 mas from the core, then turned sharply to the west and
then the jet went to the North-East with the direction almost
parallel to the first 5 mas jet. The low emission showed jet
turned to the West again.

%

\begin{figure*}
\vspace*{7.5cm} \includegraphics{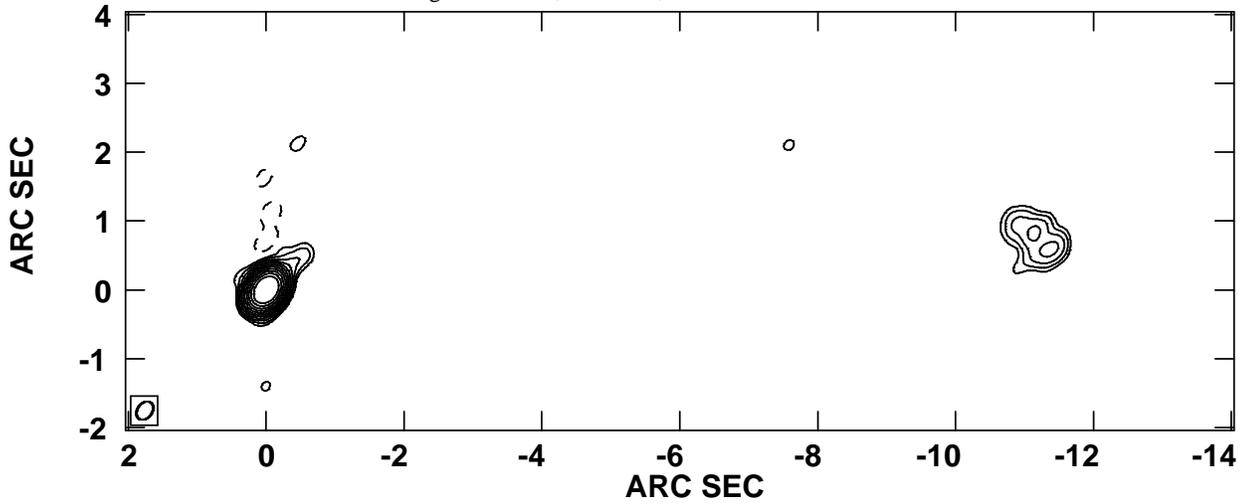} \vspace*{-1.2cm} \caption{A
5~GHz MERLIN image of NRAO530 at the epoch 1999.12,
 peak flux density= 4.55~Jy/beam, the r.m.s. noise is 0.2~mJy/beam,
 the lowest contour is 1.0 mJy/beam, contour levels increase by a factor
 of 2, FWHM: $0.29\arcsec\times0.27\arcsec$ ; P.A.$-36.7\degr$.}
\end{figure*}

\begin{figure*}
\vspace*{8cm} \includegraphics{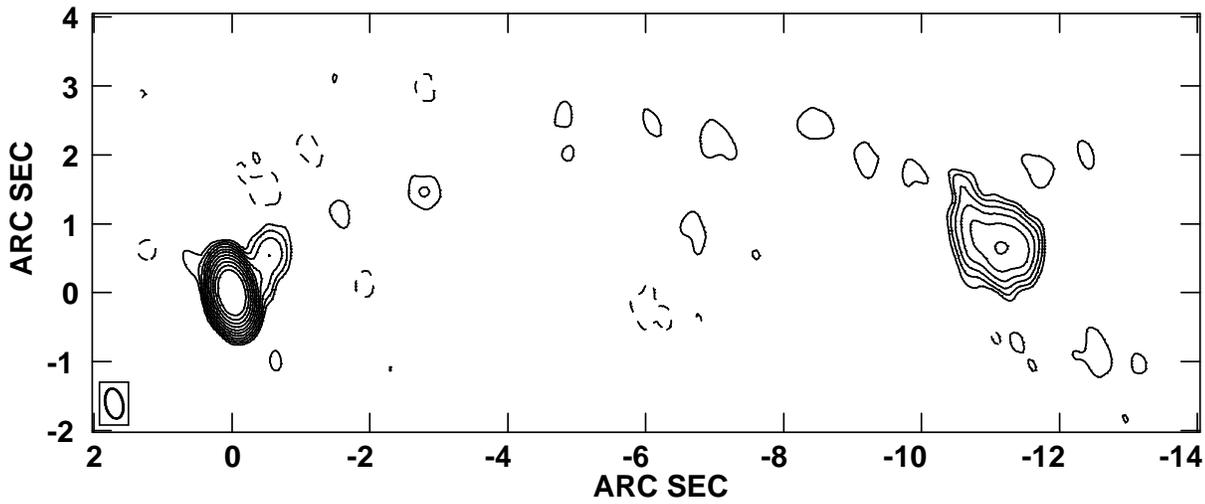} \vspace*{-1.2cm} \caption{A
1.6~GHz MERLIN image of NRAO530 at the epoch 1998.35,
 peak flux density= 5.03~Jy/beam, the r.m.s. noise is 0.3~mJy/beam,
 the lowest contour is 1.0 mJy/beam, contour levels increase by a factor
 of 2, FWHM: $0.44\arcsec\times0.25\arcsec$; P.A. $12.1\degr$.}
\end{figure*}

\begin{figure*}
\vspace*{8cm} \includegraphics{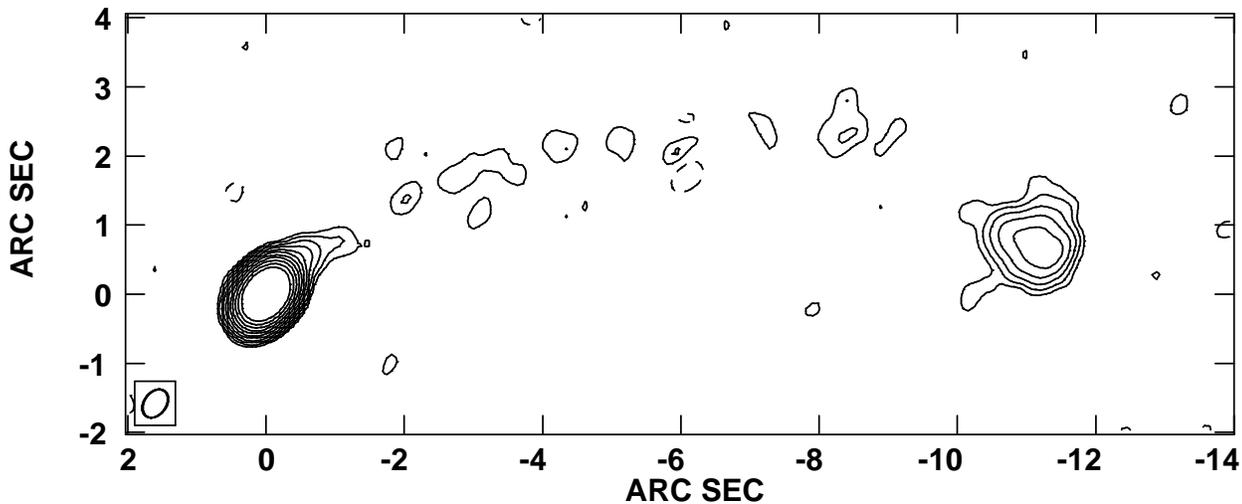} \vspace*{-1.2cm} \caption{A 8.15~GHz VLA
image of NRAO530 at the epoch 1992.9,
 peak flux density= 5.58~Jy/beam, the r.m.s. noise is 0.15~mJy/beam,
 the lowest contour is 0.45 mJy/beam, contour levels increase by a factor
 of 2, FWHM: $0.45\arcsec\times0.32\arcsec$; P.A. $-36\degr$.}
\end{figure*}

\begin{figure*}
\vspace*{9.5cm} \includegraphics{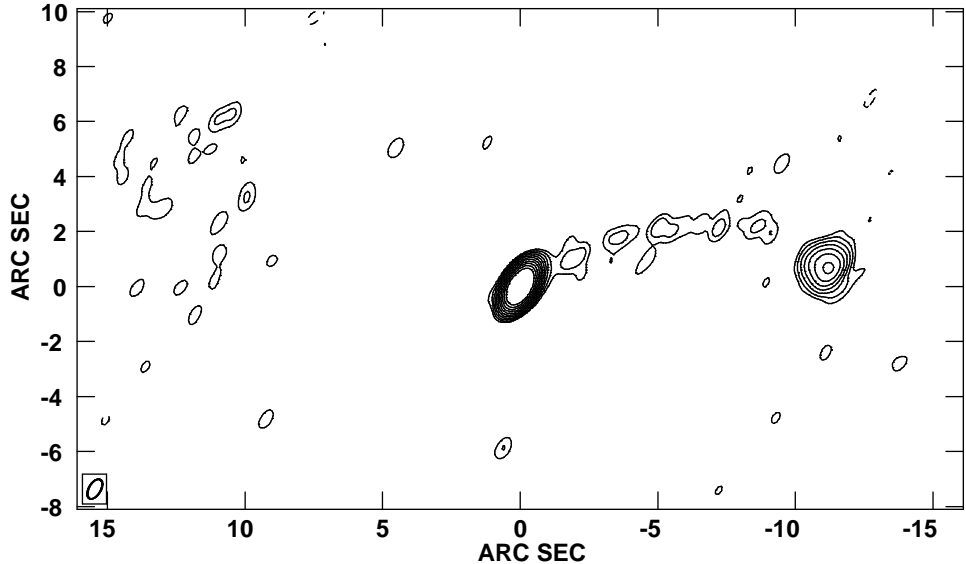} \vspace*{-2. cm} \caption{A
5~GHz VLA image of NRAO530 at the epoch 2003,
 peak flux density= 4.38~Jy/beam, the r.m.s. noise is 0.15~mJy/beam,
 the lowest contour is 0.4 mJy/beam, contour levels increase by a factor
 of 2, FWHM: $0.79\arcsec\times0.43 \arcsec$; P.A. $-29.9\degr$.}
\end{figure*}


%

\subsection{MERLIN images}

To obtain the extended emission of the source, we tapered the
5~GHz MERLIN data at 1 mega-wavelengths with a comparable
resolution of the MERLIN data at 1.6~GHz. The 5~GHz MERLIN image
shows a core-jet structure in P.A. $\sim -50\degr $ on the scale
of about 1$\arcsec$ and a distant component located at about
11$\arcsec$ west to the core. The west distant component extended
in north-east to south-west direction and resolved into two
components at resolution of $291\times272$ mas, $-36.7\degr$. No
clear connection between the core and the west distant component
was detected in the MERLIN observation (Fig. 2).

The 1.6~GHz MERLIN image shows a similar morphology to the 5~GHz
MERLIN image. The core-jet structure and the west component are
clearly detectable. Some week emission with an arch shape
connecting between the core and the distant component has been
detected in 1.6~GHz MERLIN observation (Fig. 3). This indicates
that the west component is the lobe or hotspot of NRAO530.

\subsection{VLA images}

The VLA images at 8.1 and 4.8~GHz (Figs. 4\&5) confirm the
detection of the low emission between the core and the west
distant component (let's call it the west lobe). Both images
exhibit the same track, like a ballistic trajectory, from the core
to the west lobe. This suggests that the emitting materials in the
western lobe is powered by the nuclear source.

A patch of weak diffuse emission is detected in the region about
12.5$\arcsec$ nearly east (PA=70 $\degr$) to the core, which
appears to be a counter lobe of the source. No clear connection
between the east lobe and the core has been detected. However, the
orientation of the milli-arcsec jet suggests that the lobe could
be also powered by the nuclear source.

\begin{figure}
\vspace*{6.5cm} \includegraphics{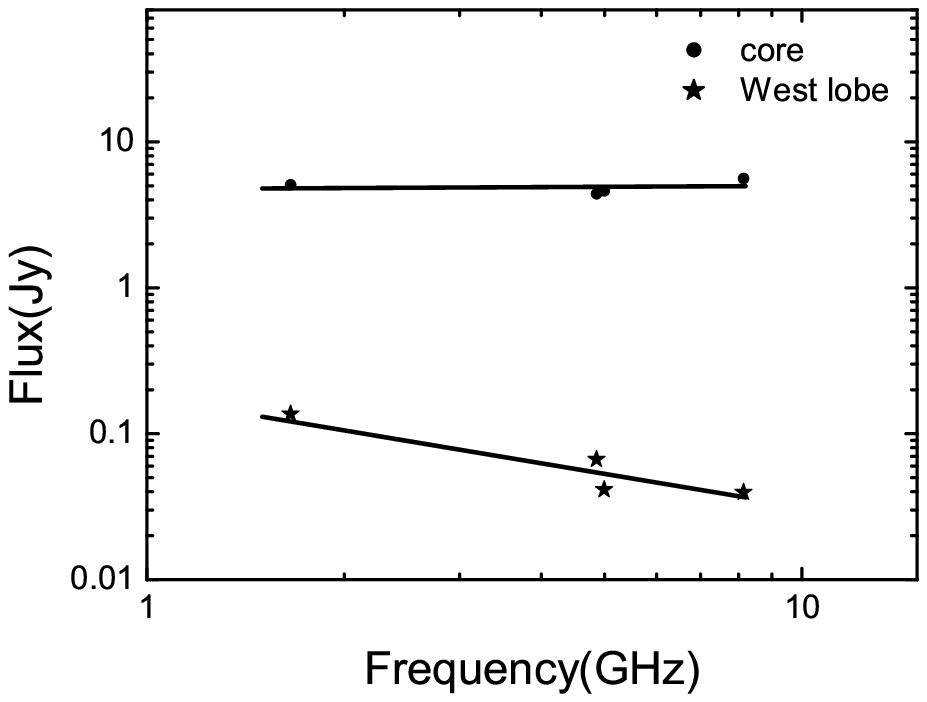} \vspace*{-.5cm} \caption{ The
spectral indexes of the core and the west lobe.}
\end{figure}

\section{Conclusions and discussion}

On the scale of parsec, NRAO530 appears to show that an oscillating
jet consists of a number of emission components
north to the core.

On the scale of kilo-parsec, the source exhibits a core jet
structure in P. A. $-50\degr$ and double lobe in the East-West
direction. The west lobe is much stronger than the east lobe with
the ratio about 5.5 in intensity.  The core showed a flat spectral
index ($\alpha=-0.02, S\propto\nu^{-\alpha}$), while the west lobe
showed a steep spectral index ($\alpha=0.8$). Fig. 6 shows the
spectrum of the core component and the west lobe with the four
epochs MERLIN and VLA data. Table 1 gives the spectral indexes of
the core component and the west lobe. A low emission are clear
detected between the core component and the west lobe, while the
connection between the core and the east lobe has not been
detected.

It is not clear why the jet bents almost 90$\degr$ from pc scale
to kpc scale, which is not common for EGRET-detected AGNs (Hong et
al. 1998). It may be caused by the interaction between the jets
and the interstellar medium (IMS) of the host galaxy. It is also
possible that the jet moves in a helical trajectory as 1156+295
(Hong et al. 2004).

\begin{acknowledgements}

This research is supported by the National Science Foundation of
PR China (10328306, 10333020, and 10373019). The authors are
grateful for the technique support by the staff of the EVN, MERLIN
and VLA. The European VLBI Network is a joint facility of
European, Chinese, South African and other radio astronomy
institutes funded by their national research councils. MERLIN is a
National Facility operated by the University of Manchester at
Jodrell Bank Observatory on behalf of PPARC. The VLA is a facility
of the National Radio Astronomy Observatory, which is operated by
Associated Universities Inc. under cooperative agreement with the
National Science Foundation.  X.Y.Hong thanks the MERLIN staff Dr.
S. T. Garrington for his kind help for the calibration of the
MERLIN data and JIVE staff Dr. L. I. Gurvits for his useful
comments. This research has made use of the NASA/IPAC
Extragalactic Database (NED).

\end{acknowledgements}

\end{document}